\documentclass[preprint,aps,showpacs,nofootinbib,preprintnumbers,amsmath,amssymb]{revtex4-1}
\usepackage{epsfig}
\usepackage{subfigure}
\usepackage{dcolumn}
\usepackage{bm}
\usepackage[usenames ,dvipsnames]{xcolor}
\usepackage{slashed}
\usepackage{graphicx,color}

\begin{document}
\title{Three-body charmless baryonic $\bar B^0_s$ decays}

\author{C.Q. Geng$^{1}$, Y.K. Hsiao$^{1}$ and Eduardo Rodrigues$^2$}
\affiliation{
$^{1}$Chongqing University of Posts \& Telecommunications, Chongqing,  400065, China\\
Department of Physics, National Tsing Hua University, Hsinchu, Taiwan 300\\
$^{2}$Department of Physics, University of Cincinnati, Cincinnati, Ohio 45221, USA
}
\date{\today}

\begin{abstract}
We study for the first time the three-body charmless baryonic decays
$\bar B^0_s\to \bar p \Lambda M^+ (p \bar \Lambda M^-)$,
with $M=\pi, K$.
We find that the branching ratios of $\bar B^0_s\to (\bar p \Lambda K^+$
and $p\bar \Lambda K^-)$ and $\bar B^0_s\to p\bar \Lambda \pi^-$
are $(5.1\pm 1.1)\times 10^{-6}$ and $(2.8\pm 1.5)\times 10^{-7}$, respectively,
which agree with recent experimental results reported by the LHCb collaboration.
In addition, we derive the relations
\mbox{${\cal B}(\bar B^0_s\to \bar p \Lambda K^+)\simeq 
(f_K/f_\pi)^2(\tau_{B^0_s}/\tau_{B^0})
{\cal B}(\bar B^0\to \bar p \Lambda \pi^+)$} and 
${\cal B}(\bar B^0_s\to p\bar \Lambda \pi^-)/{\cal B}(\bar B^0_s\to p\bar \Lambda K^-)
\simeq {\cal B}(B^-\to p\bar p \pi^-)/{\cal B}(B^-\to p\bar p K^-)$
to be confronted to future experimental measurements.
The fact that all four processes
$B^0_s, \bar B^0_s \to p\bar \Lambda K^-, \bar p \Lambda K^+$ can occur
opens the possibility of decay-time-dependent CP violation measurements
in baryonic $B$ decays, something that had not been realised before.
\end{abstract}

\maketitle
\section{introduction}
In contrast with mesonic $B$ decays,
the decays of $B$ mesons to baryonic final states have been observed to have 
unique signatures due to the baryon-pair $({\bf B_1\bar B_2})$ formations,
which reflect rich mechanisms for the hadronizations of the spinors.
For example, the BaBar and Belle experiments at the $B$ factories~\cite{pdg}
reported typical three-body charmless baryonic $B$ decay branching ratios
${\cal B}(B\to {\bf B_1\bar B_2}M)\simeq {\cal O}(10^{-6})$,
and provided evidence for prominent peaks around
$m_{\bf B_1\bar B_2}\simeq m_{\bf B_1}+m_{\bf \bar B_2}$ in the baryon-antibaryon
spectra of baryonic $B$ decays~\cite{Abe:2002ds}, 
which show that the $\bf B_1\bar B_2$ formations favour the threshold area.
However,
in two-body decays $B\to {\bf B_1\bar B_2}$,
there is no large energy release from the recoiled meson~\cite{Hou:2000bz},
such that the total energy of ${\bf B_1\bar B_2}$ is at the $m_B$ scale, 
which definitely deviates from the threshold area~\cite{Hsiao:2014zza}.
As a result, ${\cal B}(B\to {\bf B_1\bar B_2})$ are seen to be small,
around $10^{-8}-10^{-7}$~\cite{Aaij:2013fta,Aaij:2014tua,Aaij:2016xfa}.
Furthermore, the angular distribution asymmetry ${\cal A}_\theta$
of $\bar B^0\to \bar p \Lambda \pi^+$ has been measured to have an
unexpectedly large value of $(-41\pm 11\pm 3)\%$,
indicating significant interference as a result of 
the baryonic form factors~\cite{Geng:2006wz,Hsiao:2016amt}.
The same behaviour has been observed in decays to final states with open charm,
for example ${\cal A}_\theta(\bar B^0\to \Lambda\bar p D^{*+})
=(55\pm 17)\%$~\cite{Chang:2015fja}.

The aforementioned observations in $\bar B^0/B^-\to {\bf B_1\bar B_2}(M)$ decays
may also hold for $\bar B^0_s\to {\bf B_1\bar B_2}(M)$ decays
now experimentally accessible to the LHCb
collaboration~\cite{Aaij:2013yba,Hsiao:2014tda}.
Nonetheless,
baryonic $\bar B^0_s$ decays are not trivially related
to baryonic $\bar B^0$ and $B^-$ decays.
For example, 
replacing $(\bar u,\bar d)$ by $\bar s$ in $\bar B^0/B^-$,
one may approximately infer that
\begin{eqnarray}\label{relation1}
{\cal B}(\bar B^0_s\to \bar p \Lambda K^+)&\simeq&{\cal B}(\bar B^0\to \bar p \Lambda \pi^+)\,,\nonumber\\
{\cal B}(\bar B^0_s\to p\bar \Lambda \pi^-)&\simeq&{\cal B}(B^-\to p\bar p \pi^-)\,,\nonumber\\
{\cal B}(\bar B^0_s\to p\bar \Lambda K^-)&\simeq&{\cal B}(B^-\to p\bar p K^-)\,,
\end{eqnarray}
which will be shown to be mostly incorrect, except for the first relation.
We will also demonstrate that the recent first observation, made by the
LHCb collaboration, of a baryonic
$\bar B^0_s$ decay, namely $\bar B^0_s\to p\bar \Lambda K^-$,
and the measurement of its branching ratio~\cite{BsBBM},
combines in reality the branching ratios of 
$\bar B^0_s\to p\bar \Lambda K^-$ and $\bar B^0_s\to \bar p \Lambda K^+$.

\section{Formalism}
\begin{figure}[bth!]
\centering
\includegraphics[width=2.3in]{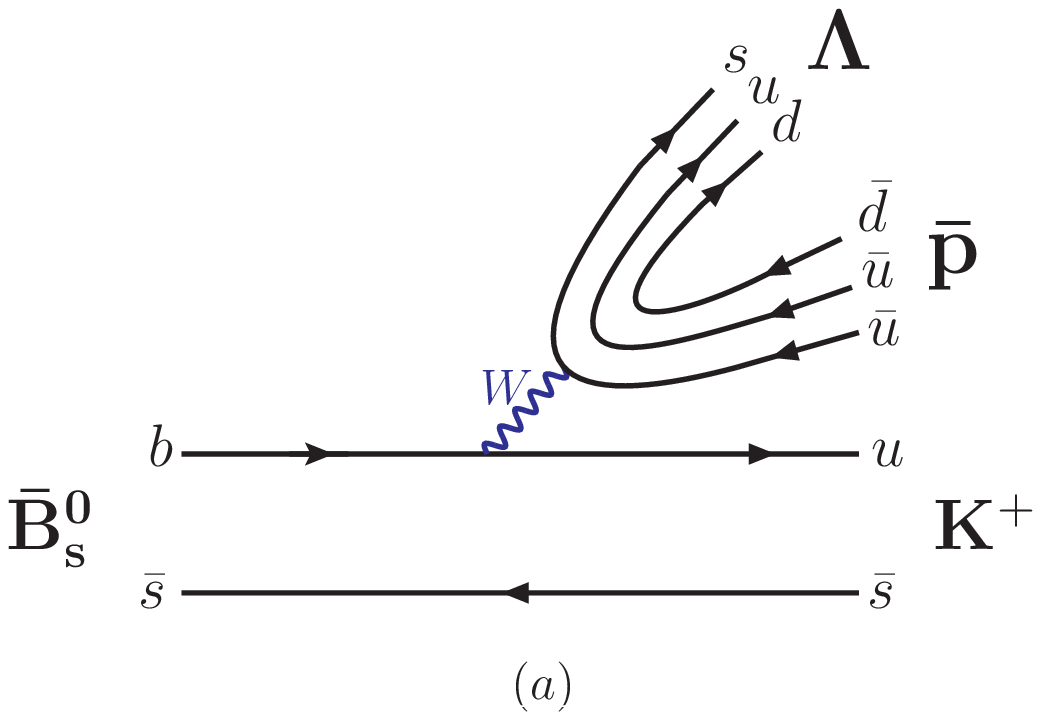}
\includegraphics[width=2.2in]{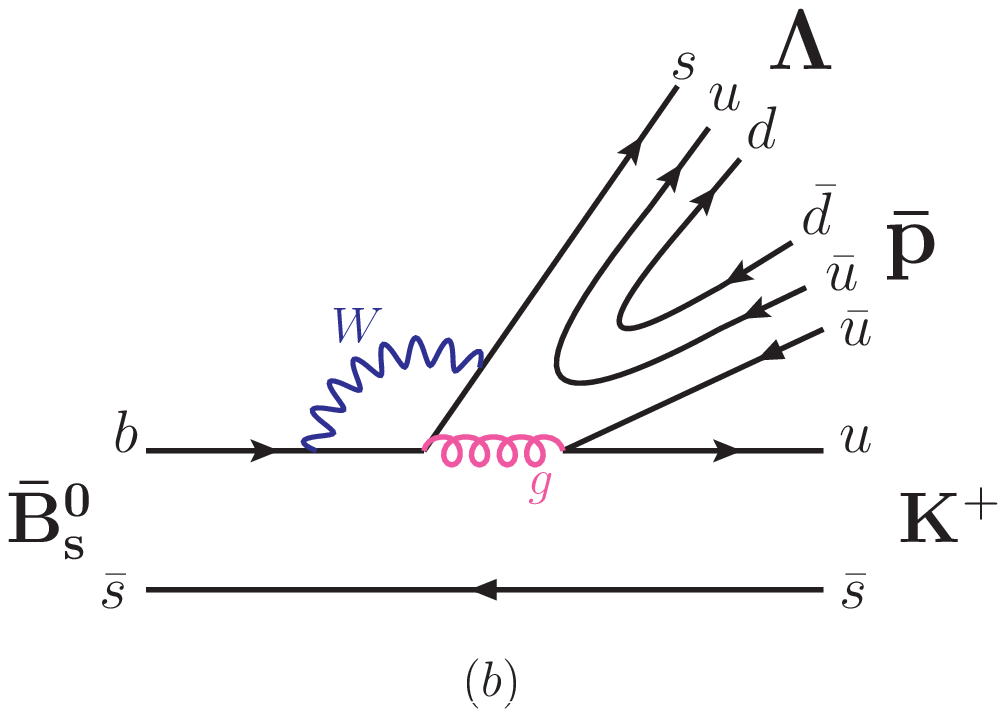}
\includegraphics[width=2.3in]{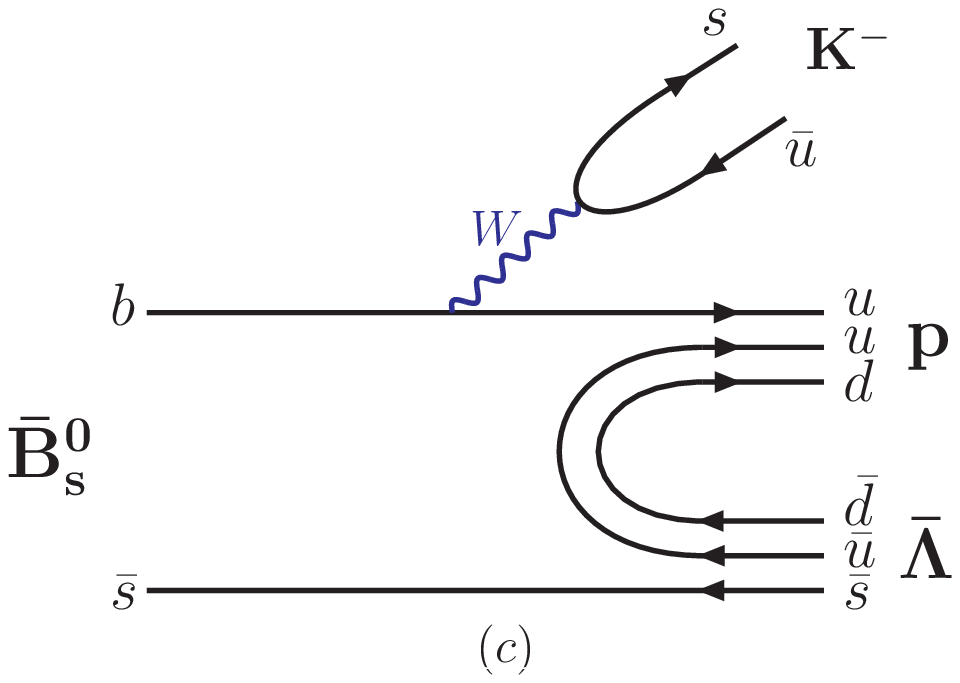}
\includegraphics[width=2.2in]{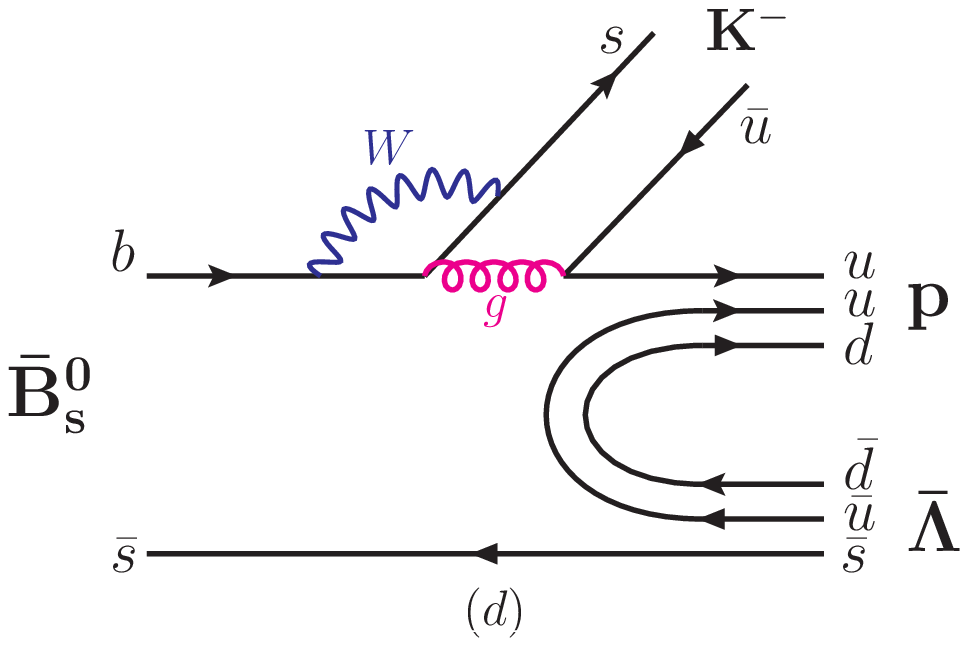}
\caption{Feynman diagrams for three-body baryonic $\bar B^0_s$ decays, where
(a,b) depict $\bar B^0_s\to \bar p \Lambda K^+$ while
(c,d) depict $\bar B^0_s\to p\bar \Lambda K^-$.}\label{dia}
\end{figure}

The decay $\bar B^0 \to \bar p \Lambda \pi^+$ is flavour specific,
unlike the similar mode of the $\bar B^0_s$ meson, which can decay to both
$\bar p \Lambda K^+$ and $p\bar \Lambda K^-$ final states.
The latter three-body baryonic $\bar B^0_s$ decays
proceed through different configurations as demonstrated
in the Feynman diagrams in Fig.~\ref{dia}.
Specifically, the baryon pairs involve quark currents
and $B$ meson transitions as depicted in Figs.~\ref{dia}(a,b) and (c,d),
respectively.

\vspace*{0.3cm}
\vspace*{0.3cm}

The amplitudes can be factorized in terms of the effective Hamiltonian
at the quark level~\cite{ali}
as~\cite{Geng:2006wz,Chua:2002wn,Chua:2002yd,Geng:2005wt,Geng:2006jt}
\begin{eqnarray}\label{amp1}
&&{\cal A}(\bar B^0_s\to \bar p \Lambda K^+)=\nonumber\\
&&\frac{G_F}{\sqrt 2}\bigg\{ \alpha_1\langle \bar p \Lambda|(\bar s u)_{V-A}|0\rangle 
  \langle K^{+}|(\bar u b)_{V-A}|\bar B^0_s\rangle
+\alpha_6\langle \bar p \Lambda|(\bar s u)_{S+P}|0\rangle
\langle K^{+}|(\bar u b)_{S-P}|\bar B^0_s\rangle\bigg\},\nonumber\\
&&{\cal A}(\bar B^0_s\to p\bar\Lambda K^-)=\nonumber\\
&&\frac{G_F}{\sqrt 2}\bigg\{ \alpha_1\langle K^{-}|(\bar s u)_{V-A}|0\rangle 
\langle p\bar \Lambda|(\bar u b)_{V-A}|\bar B^0_s\rangle
+\alpha_6\langle  K^{-}|(\bar s u)_{S+P}|0\rangle
\langle p\bar \Lambda|(\bar u b)_{S-P}|\bar B^0_s\rangle\bigg\},
\end{eqnarray}
with $\alpha_1=V_{ub}V_{us}^* a_1-V_{tb}V_{ts}^*a_4$ and $\alpha_6=V_{tb}V_{ts}^*2a_6$,
where $G_F$ is the Fermi constant, $V_{ij}$ are the CKM matrix elements,
$(\bar q_1 q_2)_{V(A)}$ and $(\bar q_1 q_2)_{S(P)}$ stand for 
$\bar q_1 \gamma_\mu(\gamma_5) q_2$ and $\bar q_1(\gamma_5) q_2$, respectively, and
$a_{1(4,6)}\equiv c^{eff}_{1(4,6)}+c^{eff}_{2(3,5)}/N_c^{eff}$ 
are composed of the effective Wilson coefficients $c_{i}^{eff}$ defined in Ref.~\cite{ali}
with $N_c^{eff}$ the effective colour number, ranging
between 2 and $\infty$ to account for the non-factorizable effects
in the generalized factorization approach. 
The amplitude ${\cal A}(\bar B^0_s\to p\bar \Lambda \pi^-)$ is obtained
from ${\cal A}(\bar B^0_s\to p\bar\Lambda K^-)$ of Eq.~(\ref{amp1})
replacing the strange quark by the down quark.

In our calculation, the matrix elements of 
$\bar B^0_s\to \bar p \Lambda K^+$ in Eq.~(\ref{amp1}) 
are expressed as~\cite{Chua:2002yd,Geng:2005wt} 
\begin{eqnarray}\label{FFactor1}
\langle M| \bar q \gamma^\mu b|B\rangle&=&(p_B+p_M)^\mu F_1^{BM}
+\frac{m^2_B-m^2_M}{t}q^\mu (F_0^{BM}-F_1^{BM})\,,\nonumber\\
\langle {\bf B_1\bar B_2}|\bar q_1\gamma_\mu q_2|0\rangle
&=&
\bar u\bigg[F_1\gamma_\mu+\frac{F_2}{m_{\bf B_1}+m_{\bf \bar B_2}}i\sigma_{\mu\nu}q_\mu\bigg]v\;,\nonumber\\
\langle {\bf B_1\bar B_2}|\bar q_1\gamma_\mu \gamma_5 q_2|0\rangle
&=&\bar u\bigg[g_A\gamma_\mu+\frac{h_A}{m_{\bf B_1}+m_{\bf \bar B_2}}q_\mu\bigg]\gamma_5 v\,,\nonumber\\
\langle {\bf B_1\bar B_2}|\bar q_1 q_2|0\rangle &=&f_S\bar uv\;,
\langle {\bf B_1\bar B_2}|q_1\gamma_5 q_2|0\rangle =g_P\bar u \gamma_5 v\,,
\end{eqnarray}
with $q=p_B-p_M=p_{\bf B_1}+p_{\bf\bar B_2}$, $t\equiv q^2$, $p=p_B-q$, and
$u$($v$) the (anti-)baryon spinor, where 
$F_{0,1}^{BM}$ are the form factors for the $B\to M$ transition, and
$F_{1,2}$, $g_A$, $h_A$, $f_S$, and $g_P$
the timelike baryonic form factors.
For $\bar B^0_s\to p\bar \Lambda K^-$,
besides $\langle M|\bar q_1 \gamma^\mu \gamma_5 q_2|0\rangle=-if_M p^\mu_M$
with $f_M$ the decay constant, 
the matrix elements of the $B\to{\bf B_1\bar B_2}$ transition 
are parameterized as~\cite{Chua:2002wn,Geng:2006wz}
\begin{eqnarray}\label{FFactor2}
%
\langle {\bf B_1\bar B_2}|\bar q\gamma_\mu b|B\rangle&=&
i\bar u[  g_1\gamma_{\mu}+g_2i\sigma_{\mu\nu}p^\nu +g_3 p_{\mu} 
+g_4q_\mu +g_5(p_{\bf\bar B_2}-p_{\bf B_1})_\mu]\gamma_5v\,,\nonumber\\
\langle {\bf B_1\bar B_2}|\bar q\gamma_\mu\gamma_5 b|B\rangle&=&
i\bar u[ f_1\gamma_{\mu}+f_2i\sigma_{\mu\nu}p^\nu +f_3 p_{\mu} 
+f_4q_\mu +f_5(p_{\bf\bar B_2}-p_{\bf B_1})_\mu]v\,,\nonumber\\
\langle {\bf B_1\bar B_2}|\bar q b|B\rangle&=&
i\bar u[ \bar g_1\slashed p+\bar g_2(E_{\bf \bar B_2}+E_{\bf B_1})
+\bar g_3(E_{\bf \bar B_2}-E_{\bf B_1})]\gamma_5v\,,\nonumber\\
\langle {\bf B_1}{\bf\bar B_2}|\bar q\gamma_5 b|B\rangle&=&
i\bar u[ \bar f_1\slashed p+\bar f_2(E_{\bf \bar B_2}+E_{\bf B_1})
+\bar f_3(E_{\bf \bar B_2}-E_{\bf B_1})]v\,,
\end{eqnarray}
where $g_i(f_i)$ $(i=1,2, ...,5)$ and $\bar g_j(\bar f_j)$ $(j=1,2,3)$
are the $B\to{\bf B_1\bar B_2}$ transition form factors.
The form factors in Eqs.~(\ref{FFactor1}) and (\ref{FFactor2})
are momentum dependent. Explicitly, $F_{0,1}^{BM}$ are given by~\cite{MS} 
\begin{eqnarray}\label{form2}
F^{BM}_1(t)=
\frac{F^{BM}_1(0)}{(1-\frac{t}{M_V^2})(1-\frac{\sigma_{11} t}{M_V^2}+\frac{\sigma_{12} t^2}{M_V^4})}\,,\;
F^{BM}_0(t)&=&\frac{F^{BM}_0(0)}{1-\frac{\sigma_{01} t}{M_V^2}+\frac{\sigma_{02} t^2}{M_V^4}}\,.
\end{eqnarray}
In perturbative QCD counting rules, 
the baryonic form factors depend on $1/t^n$ as 
the leading-order expansion~\cite{Brodsky:1973kr,Brodsky:2003gs,Chua:2002wn,Geng:2006wz},
given by
\begin{eqnarray}\label{timelikeF2}
&&F_1=\frac{\bar C_{F_1}}{t^2}\,,\;g_A=\frac{\bar C_{g_A}}{t^2}\,,\;
f_S=\frac{\bar C_{f_S}}{t^2}\,,\;g_P=\frac{\bar C_{g_P}}{t^2}\,,\;\nonumber\\
&&f_i=\frac{D_{f_i}}{t^3}\,,\;g_i=\frac{D_{g_i}}{t^3}\,,\;
\bar f_i=\frac{D_{\bar f_i}}{t^3}\,,\;\bar g_i=\frac{D_{\bar g_i}}{t^3}\,,
\end{eqnarray}
where $\bar C_i=C_i [\text{ln}({t}/{\Lambda_0^2})]^{-\gamma}$
with $\gamma=2.148$ and $\Lambda_0=0.3$ GeV.

\section{Numerical Results and Discussions }
For the numerical analysis, the theoretical inputs of the CKM matrix elements 
in the Wolfenstein parameterisation are given by~\cite{pdg}
\begin{eqnarray}
&&V_{ub}=A\lambda^3(\rho-i\eta),\,V_{tb}=1-A^2\lambda^4/2\,,\nonumber\\
&&V_{ud}=1-\lambda^2/2,,\,V_{td}=A\lambda^3\,,\nonumber\\
&&V_{us}=\lambda,\,V_{ts}=-A\lambda^2+A\lambda^4[1+2(\rho-i\eta)]/2,
\end{eqnarray}
with $(\lambda,\,A,\,\rho,\,\eta)=(0.225,\,0.814,\,0.120\pm 0.022,\,0.362\pm 0.013)$.
Other parameters are taken to be~\cite{MS} 
$F_{1,0}^{B_s K}(0)=0.31$, $(\sigma_{11},\sigma_{12})=(0.63,0.33)$, 
$(\sigma_{01},\sigma_{02})=(0.93,0.70)$, $M_V=5.32$~GeV,
and $(f_K,f_\pi)=(156.2\pm 0.7,130.4\pm 0.2)$ MeV~\cite{pdg}.
Theoretically,  
the $\bar B^0_s\to \bar p \Lambda K^+$ decay is related to
$\bar B^0\to n\bar p D^{*+}$, 
$\bar B^0\to \Lambda\bar p D^{(*)+}$,
$\bar B^0\to \bar p \Lambda \pi^+$, 
$B^-\to \bar p \Lambda (\pi^0,\rho^0)$, 
$\bar B^0_{(s)}\to p\bar p$, and
$B^-\to \bar p \Lambda$
through the timelike baryonic form factors, 
which can be connected by
the $SU(3)$ flavour and $SU(2)$ spin symmetries~\cite{Brodsky:1973kr,Chua:2002yd},
leading to~\cite{Hsiao:2016amt}
\begin{eqnarray}\label{fitC1}
&&
C_{F_1}=\sqrt\frac{3}{2}C_{||}\,,\;
C_{g_A}=\sqrt\frac{3}{2}(C_{||}+C_2)\,,\nonumber\\
&&
C_{f_S}=-\sqrt\frac{3}{2}\bar C_{||}\,,
C_{g_P}=-\sqrt\frac{3}{2}(\bar C_{||}+\bar C_2)\,,
\end{eqnarray}
where
\begin{eqnarray}\label{fitC2}
&&(C_{||},\,C_2)=(154.4\pm 12.1,\,19.3\pm 21.6)\;{\rm GeV}^{4}\,,\nonumber\\
&&(\bar C_{||},\,\bar C_2)=(537.6\pm 28.7,\,-342.3\pm 61.4)\;{\rm GeV}^{4}\,,
\end{eqnarray}
extracted from the data. Here,
$F_2=F_1/(t\text{ln}[t/\Lambda_0^2])$~\cite{Belitsky:2002kj} 
and $h_A=C_{h_A}/t^2$ have both been neglected. 
Note that
$C_{h_A}$ is fitted to be 
in accordance with 
${\cal B}(\bar B^0\to p\bar p)=1.47\times 10^{-8}$~\cite{Hsiao:2014zza}.
On the other hand, the $\bar B^0_s\to p\bar \Lambda K^-$ decay 
corresponds to
$\bar B^0\to p\bar p D^{(*)0}$,
$B^-\to p\bar p (K^{(*)-},\pi^-)$,  
$\bar B^0\to p\bar p \bar K^{(*)0}$, and
$B^-\to p\bar p e^-\bar \nu_e$
through the $B\to{\bf B\bar B'}$ transition form factors, 
which are related by the same symmetries~\cite{Chua:2002wn,Geng:2006wz},
given by
\begin{eqnarray}\label{fitD1}
&&
D_{g_1}=D_{f_1}=-\sqrt\frac{3}{2}D_{||}\,,\;D_{g_{4,5}}=-D_{f_{4,5}}=-\sqrt\frac{3}{2}D_{||}^{4,5}\,,\nonumber\\
&&
D_{\bar g_1}=D_{\bar f_1}=-\sqrt\frac{3}{2}D_{||}\,,\;
D_{\bar g_{2,3}}=-D_{\bar f_{2,3}}=-\sqrt\frac{3}{2}\bar D_{||}^{2,3}\,,
\end{eqnarray}
 with the vanishing form factors $(g_{2,3},f_{2,3})$ due to the derivations of
$f_M p^\mu\bar u (\sigma_{\mu\nu} p^\nu)v=0$ for $g_2(f_2)$ and 
$f_M p^\mu \bar u p_\mu v\propto m_M^2$ for $f_3(g_3)$ in the amplitudes, 
where
\begin{eqnarray}\label{fitD2}
&&
D_{||}=(45.7\pm 33.8)\;{\rm GeV}^{5}\,,
(D_{||}^4,D_{||}^5)=(6.5\pm 18.1,-147.1\pm 29.3)\;{\rm GeV}^{4}\,,\nonumber\\
&&\bar D_{||}=(35.2\pm 4.8)\;{\rm GeV}^{5}\,,\;
(\bar D_{||}^2,\bar D_{||}^3)=(-22.3\pm 10.2, 504.5\pm 32.4)\;{\rm GeV}^{4}\,.
\end{eqnarray}
The effective Wilson coefficients for the $\bar B^0_s (B^0_s)$ decays
are given by~\cite{ali}
\begin{eqnarray}
c^{eff}_1      &=& 1.168,\,\nonumber\\
c^{eff}_2      &=& -0.365\,,\nonumber\\
10^4 c^{eff}_3 &=& 241.9\pm 3.2\eta + 1.4 \rho + i(31.3\mp 1.4\eta + 3.2\rho),\,\nonumber\\
10^4 c^{eff}_4 &=& -508.7 \mp 9.6\eta - 4.2\rho+ i(-93.9 \pm 4.2\eta - 9.6\rho) ,\,\nonumber\\
10^4 c^{eff}_5 &=& 149.4\pm 3.2\eta + 1.4\rho + i(31.3\mp 1.4\eta + 3.2\rho),\,\nonumber\\
10^4 c^{eff}_6 &= &-645.5 \mp 9.6\eta- 4.2\rho + i(-93.9\pm 4.2\eta - 9.6\rho).\,
\end{eqnarray}

\begin{figure}[t!]
\centering
\includegraphics[width=2.3in]{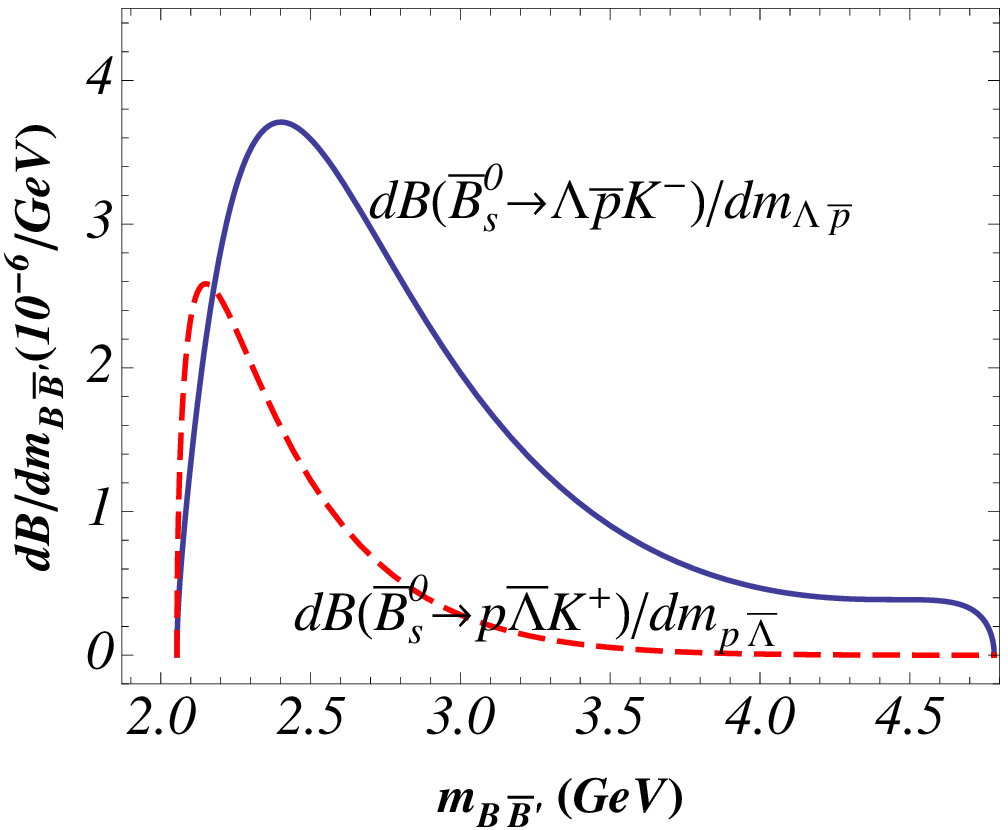}
\includegraphics[width=2.3in]{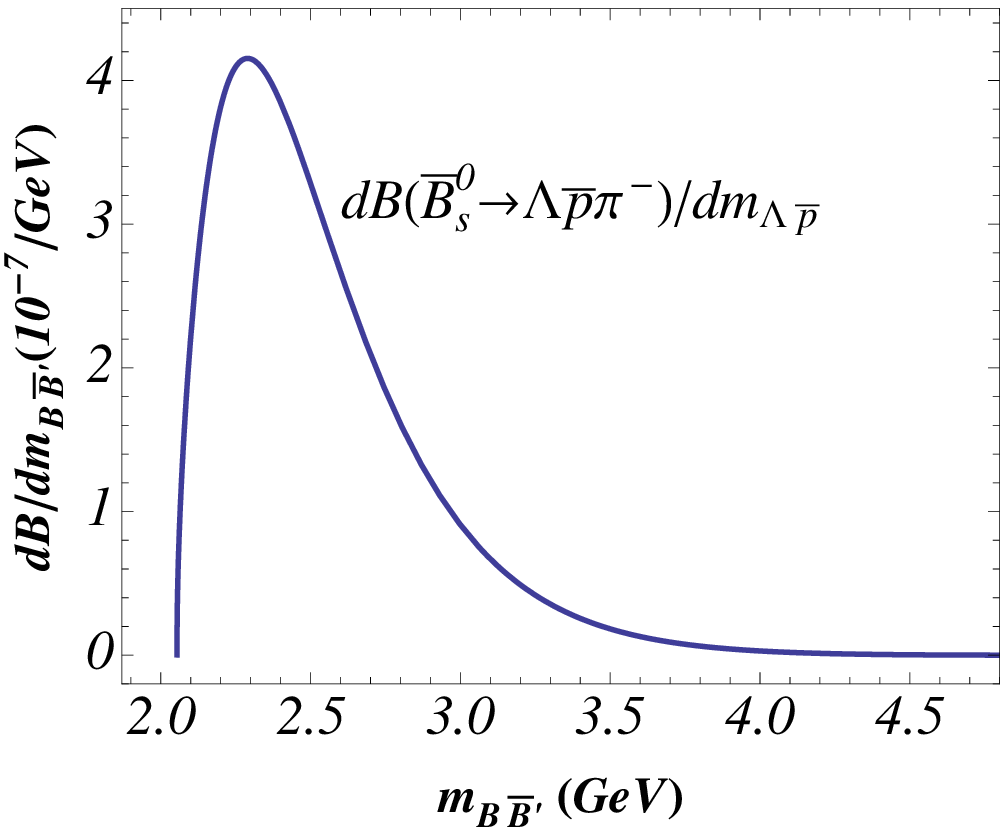}
\caption{Spectra for the three-body baryonic decays
(left) $\bar B^0_s\to (\Lambda\bar p K^+,p\bar \Lambda K^-)$ and
(right) $\bar B^0_s\to p\bar \Lambda \pi^-$.}\label{dia2}
\end{figure}
\vspace*{0.3cm}

Integrating over the phase space
of the three-body decays~\cite{pdg} we obtain the spectra for 
$\bar B^0_s\to (\bar p \Lambda K^+,p\bar \Lambda K^-)$ and 
$\bar B^0_s\to p\bar\Lambda \pi^-$ in Fig.~\ref{dia2}, 
which clearly present the threshold enhancement observed in a multitude of
baryonic $\bar B^0$ and $B^-$ decays.
The branching ratios are predicted to be
\begin{eqnarray}
{\cal B}(\bar B^0_s\to \bar p \Lambda K^+)&=&(3.75\pm 0.81^{+0.67}_{-0.31}\pm 0.01)\times 10^{-6}\,,
\nonumber\\
{\cal B}(\bar B^0_s\to p\bar \Lambda K^-)&=&(1.31\pm 0.32^{+0.22}_{-0.10}\pm 0.01)\times 10^{-6}\,,
\nonumber\\
{\cal B}(\bar B^0_s\to p\bar\Lambda \pi^-)&=&(2.79\pm 1.37^{+0.64}_{-0.30}\pm 0.17)\times 10^{-7}\,,
\label{eq:BRs}
\end{eqnarray}
with the uncertainties  from the form factors, non-factorizable effects, and CKM matrix elements
in order.
The ${\cal B}(\bar B^0_s\to \bar p \Lambda K^+)$
is calculated to be close to the observed 
\mbox{${\cal B}(\bar B^0\to \bar p \Lambda \pi^+)=(3.14\pm 0.29)\times 10^{-6}$}~\cite{pdg},
which confirms the first relation in Eq.~(\ref{relation1}). 
Nonetheless, using the experimental measurements of ${\cal B}(B^-\to p\bar p M)$ $(M=K^-,\pi^-)$~\cite{pdg},
we find that
${\cal B}(\bar B^0_s\to p\bar \Lambda M)\simeq 0.2 \times {\cal B}(B^-\to p\bar p M)$,
which disproves the other relations in Eq.~(\ref{relation1}).
The reason for this is that the $\bar B^0_s\to p\bar \Lambda$
and $B^-\to p\bar p$ transitions give different contributions.
Consequently, we should revise the relations in Eq.~(\ref{relation1}) to be
\begin{eqnarray}
{\cal B}(\bar B^0_s\to \bar p \Lambda K^+)&\simeq& 
(f_K/f_\pi)^2(\tau_{B^0_s}/\tau_{B^0})
{\cal B}(\bar B^0\to \bar p \Lambda \pi^+)\,,\nonumber\\
\frac{{\cal B}(\bar B^0_s\to p\bar \Lambda \pi^-)}{{\cal B}(\bar B^0_s\to p\bar \Lambda K^-)}&\simeq&
\frac{{\cal B}(B^-\to p\bar p \pi^-)}{{\cal B}(B^-\to p\bar p K^-)}\,.
\end{eqnarray}

From an experimental perspective, the measured branching ratio is
${\cal B}(B^0_s\to p \bar \Lambda K^- + \bar B^0_s \to p \bar \Lambda K^-)$,
given that the flavour of the reconstructed $B_s^0$ meson at production
is not determined -- the identification of the flavour at production,
a procedure known as flavour tagging, requires a decay-time-dependent analysis.
Assuming negligible CP violation,
${\cal B}(B^0_s\to p \bar \Lambda K^- + \bar B^0_s \to p \bar \Lambda K^-$)
is equivalent to the combination of the two branching ratios
${\cal B}(\bar B^0_s\to \bar p \Lambda K^+ + p \bar \Lambda K^-)=
(5.1\pm 1.1)\times 10^{-6}$. This calculation agrees well with the
experimental measurement,
${\cal B}(B^0_s\to p \bar \Lambda K^- + \bar B^0_s \to p \bar \Lambda K^-) = (5.48^{+0.82}_{-0.80}\pm 0.60\pm 0.51\pm 0.32)\times 10^{-6}$,
reported by the LHCb collaboration~\cite{BsBBM}.
In contrast, ${\cal B}(\bar B^0_s\to p\bar\Lambda \pi^-)$ 
is estimated to be of order $10^{-7}$, consistent
with its non-observation with the present data sample~\cite{BsBBM}.

All four processes
$B^0_s, \bar B^0_s \to p\bar \Lambda K^-, \bar p \Lambda K^+$ are possible,
just as in the case of the \mbox{$\bar B^0_s\to D_s^\pm K^\mp$}
decays~\cite{Aaij:2014fba}. A $B$-flavour tagged decay-time-dependent analysis
of these baryonic decay modes is necessary to disentangle
all contributions. As the ratio of the $\bar B^0_s\to \bar p \Lambda K^+$
and $B^0_s\to \bar p \Lambda K^+$ branching ratios is predicted to be
rather large, \textit{cf.} Eq.~(\ref{eq:BRs}), sizeable interference due to
$B^0_s$-$\bar B^0_s$ mixing is expected, which hints at possibly large
time-dependent CP violating asymmetries.
Time-dependent analyses require a typical minimum data sample
of order 1000 to 1500 signal candidates, see for example the LHCb analysis
presented in Ref.~\cite{Aaij:2014fba}. Extrapolating from the $260 \pm 21$
$B^0_s, \bar B^0_s \to p\bar \Lambda K^- \bar p \Lambda K^+$
candidates selected in the recent LHCb analysis~\cite{BsBBM},
assuming (as done in LHCb extrapolations) a two-fold increase in the
$b\bar{b}$ production cross-section between the first data taking period
of the LHC, and the present second period started in 2015,
we conclude that such an analysis will require the full data sample to be
collected by 2018.

Based on the observation of $\bar B^0_s\to (\bar p \Lambda K^+,p\bar \Lambda K^-)$, 
it is promising to study other charmless baryonic $\bar B^0_s$ decays such as
$\bar B^0_s\to \bar p \Lambda K^{*+},p\bar \Lambda K^{*-}$,
$\bar B^0_s\to \Lambda\bar \Lambda \phi$,
$\bar B^0_s\to (\Sigma^0\bar \Lambda,\Lambda\bar \Sigma^0,\Sigma^0\bar \Sigma^0) \phi$,
$\bar B^0_s\to \bar p \Sigma^0 K^+,p\bar \Sigma^0 K^-$, and
$\bar B^0_s\to p\bar \Sigma^0 \pi^-$.
The presence of extra resonances or neutral particles in the
final states of these decay modes makes the experimental searches
more demanding, though feasible by both the LHCb experiment and the future
Belle II experiment.

\section{Conclusions}
We have studied the three-body charmless baryonic decays
$\bar B^0_s\to \bar p \Lambda M^+$ and $p\bar \Lambda M^-$, with $M=\pi,K$.
We have predicted the combined branching ratio of 
$\bar B^0_s\to (\bar p \Lambda K^+$ and $p\bar \Lambda K^-)$
to be $(5.1\pm 1.1)\times 10^{-6}$, in good agreement with
the recently presented experimental result by the LHCb collaboration~\cite{BsBBM}.
We further obtained
${\cal B}(\bar B^0_s\to p\bar\Lambda \pi^-)=(2.8\pm 1.5)\times 10^{-7}$,
which is below the current experimental sensitivity of the LHCb analysis.
We have also presented 
useful relations between the three-body baryonic decays of 
$\bar B^0_s$ and $\bar B^0/B^-$, such as
${\cal B}(\bar B^0_s\to \bar p \Lambda K^+)\simeq (f_K/f_\pi)^2(\tau_{B^0_s}/\tau_{B^0})
{\cal B}(\bar B^0\to \bar p \Lambda \pi^+)$ and 
${\cal B}(\bar B^0_s\to p\bar \Lambda \pi^-)/{\cal B}(\bar B^0_s\to p\bar \Lambda K^-)
\simeq {\cal B}(B^-\to p\bar p \pi^-)/{\cal B}(B^-\to p\bar p K^-)$, 
which can be tested by the future experiments at LHCb.
The fact that all four processes
$B^0_s, \bar B^0_s \to p\bar \Lambda K^-, \bar p \Lambda K^+$ can occur
opens the possibility of decay-time-dependent CP violation measurements
in baryonic decays, something that had not been realised before.

\section*{ACKNOWLEDGMENTS}
The work of C.Q.~Geng and Y.K.~Hsiao was supported, in part, by National Center for Theoretical Sciences, 
MoST (MoST-104-2112-M-007-003-MY3), and 
National Science Foundation of China (11675030).
The work of E. Rodrigues was supported, in part,
by U. S. National Science Foundation award ACI-1450319.

\end{document}